\documentclass{elsart}
\journal{Astroparticle Physics}
\usepackage{natbib}
\usepackage{graphicx}
\usepackage{amssymb}
\usepackage{amsmath}
\usepackage{bm}

\def\url#1{{\ttfamily\def\/{/\discretionary{}{}{}}#1}}
\def\bibcode#1{}

\begin{document}
\begin{frontmatter}
\title{Color, 3D simulated images with shapelets}

\author[UCB,JPL,Caltech]{Matt Ferry\corauthref{cor}},
\corauth[cor]{Corresponding author.}
\ead{mferry@astro.caltech.edu}
\author[JPL,Caltech]{Jason Rhodes},
\author[ROE]{Richard Massey},
\author[UCB]{Martin White},
\author[JPL,Caltech]{Dan Coe},
\author[UCR]{Bahram Mobasher}

\address[UCB]{University~of~California,~Berkeley,~CA~94720}
\address[JPL]{Jet~Propulsion~Laboratory,~California~Institute~of~Technology,~Pasadena,~CA~91109}
\address[Caltech]{California~Institute~of~Technology,~1200~E.~California~Blvd.,~Pasadena,~CA~91125}
\address[ROE]{Royal~Observatory,~Blackford~Hill,~Edinburgh~EH9~3HJ,~UK}
\address[UCR]{University~of~California,~Riverside,~CA~92521}

\begin{abstract}

We present a method to simulate color, 3-dimensional images taken with a space-based observatory by building
off of the established {\em shapelets} pipeline.  The simulated galaxies exhibit complex morphologies, which
are realistically correlated between, and include, known redshifts. The simulations are created using
galaxies from the 4 optical and near-infrared bands (\textit{B}, \textit{V}, \textit{i} and
\textit{z}) of the Hubble Ultra Deep Field (UDF) as a basis set to model morphologies and redshift.  We
include observational effects such as sky noise and pixelization and can add astronomical signals of
interest such as weak gravitational lensing.  The realism of the simulations is demonstrated by comparing
their morphologies to the original UDF galaxies and by comparing their distribution of ellipticities as a
function of redshift and magnitude to wider HST COSMOS data.  These simulations have already been useful for
calibrating multicolor image analysis techniques and for better optimizing the design of proposed space
telescopes.  

\end{abstract}

\end{frontmatter}

\newpage 

\section{Introduction}

As astronomical surveys become deeper and wider, analysis techniques correspondingly become more complex and
demanding. To calibrate these methods, extensive work has already been invested in the simulation of
monochromatic astronomical imaging. Simulation packages have been developed to incorporate a semi-analytic
model of galaxy number counts and evolution \cite{EWBM}, or to mimic the properties of real observations
\cite{MRCB}. However, there are currently no packages able to create correlated images across several
bands.

Multi-band image simulations are firstly useful to develop and calibrate analysis methods that use 
multicolor data. Many measurements in astronomy  (for example photometry, astrometry and shape measurement)
are  ``inverse problems,'' where variation in a signal is easy to introduce but difficult to measure,
usually due to complications involving observational seeing and noise. Simulated data provide the best
way to calibrate such methods because these variables can be controlled. A known astronomical signal can
be inserted into simulated data, and the accuracy of a method can be judged by examining any errors in its
recovery.

One example is the measurement of weak gravitational lensing. 
In weak lensing, light from background galaxies is lensed by
foreground matter distributions, causing a shear (distortion) of the background
galaxies' shapes. The distortion is easy to add during the construction of simulated data.
Although the lensing signal is achromatic, color simulations can be used to test
sophisticated measurement methods that take advantage of
\begin{itemize}
\item the increased number of shear-measurable galaxies if some galaxies 
are only sufficiently bright in certain bands,
\item reduced {\em noise} on shear measurement  (by $\sqrt{N}$) if the intrinsic shapes of galaxies
are uncorrelated between $N$ bands, and 
\item reduced systematic {\em bias} on shear measurement if the intrinsic shapes of galaxies are correlated 
between bands \cite{JainJarvis}. 
\end{itemize}
One common challenge in weak lensing measurement is the deconvolution of galaxy shapes from the instrumental
point-spread function (PSF). Since the PSF is different in each band, PSF-dependent biases will be averaged
out by looking at multiple  bands. Conversely, biases inherent to a method will not be ameliorated.
Developing multicolor analysis techniques to exploit these tricks requires multicolor simulations.

Multi-band image simulations are also useful to optimize the design and improve the science case for
planned, multi-band imaging surveys such as SNAP \cite{Aldering} or Euclid \cite{DUNE}.  These surveys
require multiple bands in order to observe different types of galaxies, observe objects typically obscured
in other bands, observe objects out to different redshifts, and most importantly to obtain photometric
redshifts for galaxies. Engineering requirements for the design of these instruments can be derived via
image simulations by measuring the (often complex and subtle) effects of engineering parameters on
scientific return \cite{High}. Predictions for the scientific return of a given mission can be similarly
estimated \cite{WLII,WLIII}.

A full demonstration of the potential gains in multicolor shear measurement, or a full optimization of a
future space-based lensing mission is beyond the scope of this paper. The purpose of this paper is to
present a method for simulating deep, multi-color space-based images with correlated morphologies and
redshifts. The simulation pipeline we present here will serve as a basis for performing the optimization of
both shear measurement techniques and future space missions in future papers. 

Our simulation pipeline generalizes the single-color method of \cite{MRCB}, representing complex galaxy
morphologies as ``shapelets'' \cite{Cartesian,Polar}.
Shapelet-based simulations are already widely used for weak lensing.
The Shear TEsting Program (STEP) used similar simulated data to test and
improve shape measurement and PSF correction methods \cite{Mass2006}. 
Our generalization to multi-band, 3-dimensional simulations thus increases the realism and utility 
of a well established technique.  

This paper is organized as follows.  In \S 2 we give a brief review of shapelets and how they can be used to
generate simulated images.  In \S 3 we present the methodology by which we create multi-band, 3-dimensional
simulations.  In \S 4 we test the realism of our simulations through comparison to the real HST data. 
Lastly, in \S 5, we discuss the conclusions and summarize our findings.

\section{Background}\label{sec:back}

Shapelets, or 2-dimensional Gaussian-weighted Laguerre polynomials,
form a complete, orthonormal basis able to represent any localized image, including a galaxy shape, in
a relatively small number of coefficients. 
Any image $f(\underline{x})$ can be represented as a linear combination of 
shapelet basis functions $\chi_{nm}(\underline{x};\beta)$:

\begin{equation}\label{eq:shape}
f(\underline{x}) = \sum_n \sum_m f_{nm} \chi_{nm} (\underline{x};\beta),
\end{equation}

\noindent where $\underline{x}$ is the pixel position, $f_{nm}$ are the shapelet coefficients, and $\beta$
is a characteristic size \cite{Polar}.  Shapelets also simplify the practical processes of image convolution
and deconvolution.  In real space, convolution is an expensive process with computation time scaling with the
square of the number of pixels.  With shapelets, convolution becomes a computationally inexpensive matrix
operation \cite{Cartesian}, and deconvolution merely requires a matrix inversion \cite{BR}.  This is
advantageous when it is necessary to deconvolve with a PSF and re-convolve with a different PSF.

Shapelet coefficients thus form a multi-dimensional parameter space that describes a galaxy. In general, any
possible galaxy morphology can be thought of as a point in this multi-dimensional parameter space. When
the shapelet coefficients of a set of observed galaxies are placed in this space, various correlations
emerge.  Different directions in parameter space correspond to characteristics of the galaxy such as size, ellipticity, or
the number of spiral arms.  A classic example of this effect is the Hubble tuning-fork diagram
\cite{Hubble,Sandage,deV}, which parametrizes galaxies' ellipsoid, bulge/disk ratio, and how tightly
wound the spiral arms are.  The shapelets method increases the dimensionality of the parametrization with
axes corresponding to galaxies' magnitude, size ($\beta$), and polar shapelet coefficients \cite{MRCB}.

Real galaxy morphologies only occupy a small region of this multi-dimensional parameter space. Most regions
of parameter space, corresponding to random shapelet coefficients, do not produce an image that resembles a
galaxy. To manufacture useful simulations, it is essential to map the region corresponding to
morphologically realistic galaxies. This region will constitute a probability density function (PDF), from
which we will be able to draw simulated galaxy images. To acquire the PDF, we begin with a sample of real
galaxies. Of course, the PDF is only noisily sampled by this finite set of galaxies, so we smooth it to
obtain an approximation to the true, underlying PDF.

\section{Methodology}

In this section we present the methodology used to create the
simulations. The process can be summarized in two steps: (1) shapelet catalog creation, and (2) image constitution.

\subsection{Shapelet Catalog Creation}

Since our goal is to simulate multicolor images, it is necessary to start from real, multicolor data.  The
best data for this are the Hubble Ultra-Deep Field (UDF) images. In this field, there are 8049 galaxies with a
detection signal to noise ratio of at least 10 in any one band. Photometric redshifts for each galaxy are
publicly available in the Coe \textit{et al} \cite{photo-z} catalog.\footnote{Available for download at
\underline{http://adcam.pha.jhu.edu/$\sim$coe/UDF/}.}

We use the program \textit{shex} from the Shapelets software package\footnote{Version 2.1$\beta$, available
at \underline{http://www.astro.caltech.edu/$\sim$rjm/shapelets/}.} to decompose all of these galaxies, in all
observed bands, into a linear combination of shapelet basis functions.  We run \textit{shex} up to a maximum
radial oscillation value, or \textsc{n\_max} in the basis functions, of 20 in order to optimize
decomposition.  Although this is computationally expensive (\textsc{n\_max} = 20 corresponds to 231
coefficients), we are assured that the large objects are well modeled. This algorithm automatically copes
with the varying pixel scale between optical and near-infrared imaging.
To maximize the efficiency of the shapelet model, we iterate 
the center of each decomposition on the pixel grid, the maximum order $n_{\rm max}$, 
and the scale size $\beta$ of each decomposition independently in each band, using the algorithm
discussed in \cite{Polar}. This number is recorded for later image reconstruction. We store catalogs of galaxy shapes in their raw form as well as deconvolved from the UDF PSF as modeled by the stars in the field.

To model $n$-band imaging, we thus increase the dimensionality of the shapelet parameter space
$n$-fold. For example, while a bright object may be uniquely described in one band by 233 coefficients
(including magnitude, size and redshift) if \textsc{n\_max}=20, it is now described by 932 coefficients.  Though in the UDF $n=4$, our simulation software is not limited to this number
and could accept an input catalog of galaxies with more bands in the future. The UDF galaxies in this highly
dimensional parameter space automatically contain the correlations between shapelet coefficients necessary
to produce realistic galaxy images.

We then smooth the finite number of points in shapelet parameter space, using an \textit{Epanechnikov}
kernel \cite{Epanech} with a different smoothing length, $\lambda_i$, for each parameter. Note that, if we
choose $\lambda_i$ to be too small, the galaxy appears nearly unchanged, and we shall simply reproduce UDF
galaxies in the simulated images; if we choose it to be too large, the galaxy is not realistic.  Following the
established smoothing scheme explored by \cite{MRCB}, we smooth the complex polar shapelet coefficients in
modulus and phase space, setting $\lambda_i=15^{\circ}$ for phases and the mean separation between nearest
neighbors in that dimension for moduli.  To perturb the galaxy redshifts slightly, we
set the redshift smoothing length to be $\frac{1+z}{m}$, where $m$ is a free parameter.  We choose to smooth
over $1+z$ since it is used more frequently in determining cosmological parameters.  We also choose $m$ to
be 6 as a reasonable limit to conservative smoothing.  If $m$ is chosen to be lower than 6, the high
redshift objects could get smoothed to an unrealistically high redshift.

\subsection{Image Creation}\label{sec:im_create}

For each simulated image, we generate a sufficient number of new galaxies that their density in the
simulated image reproduces that in the UDF. In practice, rather than pixelating and drawing from the
smoothed PDF, we use an equivalent Monte-Carlo bootstrap technique \cite{MRCB}.  For each new galaxy, an
original UDF galaxy is selected at random and perturbed in shapelet space, within the smoothing kernel, to
create a new galaxy. We also append a mock catalog of photometric redshifts to these new galaxies.  These
redshifts are slightly perturbed from the original galaxy's observed redshift via the same smoothing
process, wile their distribution still follows the observed distribution in the UDF.

At this stage, a known weak lensing shear signal can be added to the objects. Similar effects could also be
added to simulate, e.g. proper motions, photometric variability, or supernovae. The galaxy is finally
convolved in shapelet space with the desired point spread function.

Once new objects are created, they are formed into a multi-band shape catalog to be arranged into new
images.  The objects are first re-composed into pixelated postage stamps, then placed into large, empty
arrays.  The placement of objects is done such that the object appears at the same (RA, Dec) position in
each band, in flux units of photons per second per pixel, and with (for the sake of this paper) a constant
pixel scale of $0.03$ arcsec per pixel.  Together with the mock photometric redshift catalog, a
3-dimensional, color simulation is thus created.

\begin{figure}[!h]
\begin{center}
\includegraphics[scale=0.3]{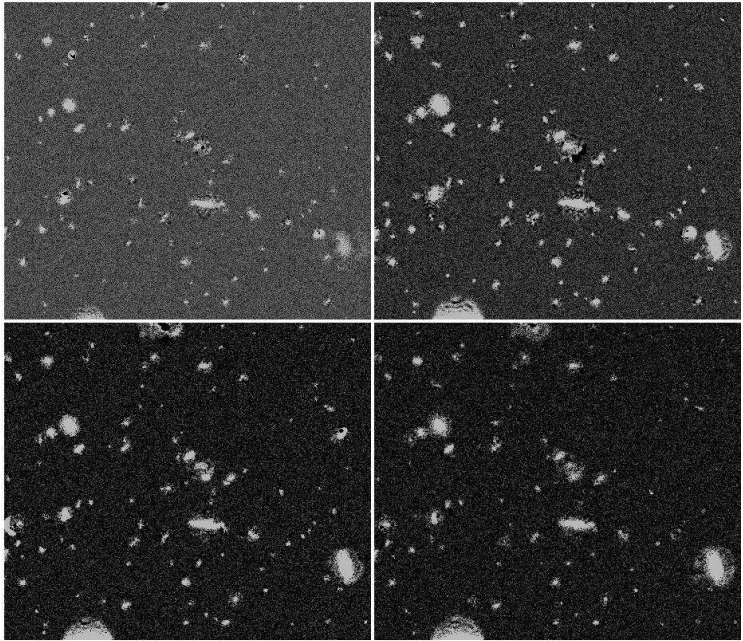}
\end{center}
\caption{Sample correlated image to UDF depth.  The bands shown are, clockwise from top-left, \textit{B} (top-left), \textit{V} (top-right), \textit{i} (bottom-left) and \textit{z} (bottom-right).  The images are displayed on the same logarithmic scale with the same contrast.}\label{fig:sample}
\end{figure}

The images are made realistic by adding both a sky background and shot noise. Figure $\ref{fig:sample}$
shows an example image. The simulations could also be made more realistic by adding cosmic rays, variable
sky background/read noise, or charge transfer inefficiency trailing. The background noise could also be
smoothed with a small kernel to approximate the effects of the \texttt{DRIZZLE} routine on stacked data from
multiple, dithered exposures \cite{Drizzle}.  On the one hand, \texttt{DRIZZLE} allows for a sharper pixel
scale and correction for geometric distortions.  On the other hand it produces correlated pixel noise. We
have not enabled these options in the standard images studied in this paper, but intend to explore their
effects in future work.

\section{Results}

We examine the morphological realism of our simulations by ``blind-testing'' them
against the original UDF.  We also create a ``$\delta$-\textit{function image},'' whereby the smoothing length ($\lambda_i$) for each coefficient is set to zero (ie, objects are not perturbed).  This allows us to separate the morphological characteristics due to shapeletization and due to smoothing in shapelet space, the results of which are found in Table $\ref{tab:morph}$.  Our tests are similar to those done in \cite{MRCB}.  We first consider general comparisons using photometry and
size from \texttt{SExtractor}.  We then
examine more advanced morphological tests.

\begin{figure}[!h]
\begin{center}
\begin{tabular}{ccc}
\includegraphics[scale=0.5]{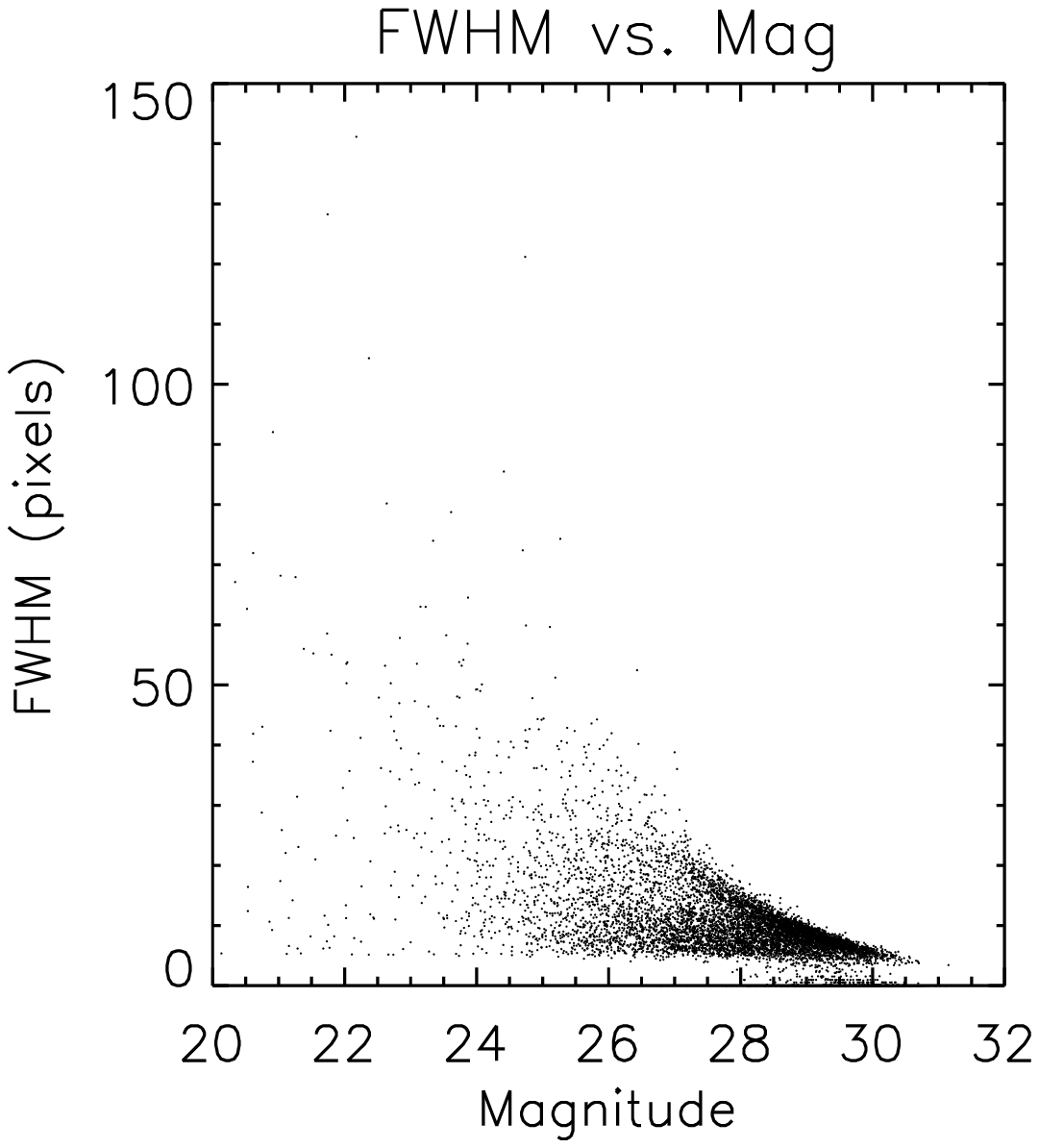}
&
\includegraphics[scale=0.5]{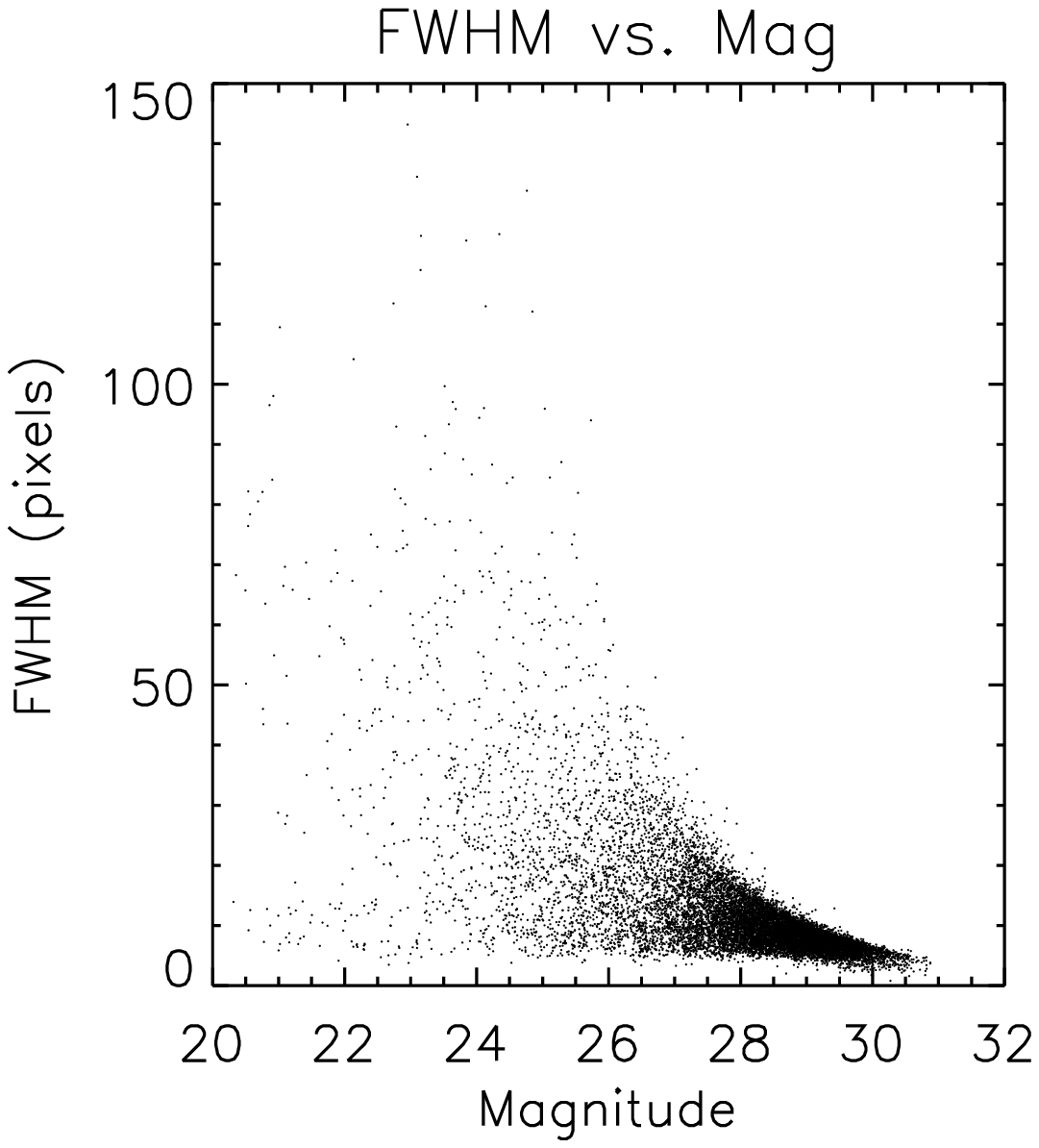}
\\
\end{tabular}
\end{center}\caption{\textsc{fwhm} vs. \textsc{mag} for i-band real UDF image
(left) and simulated image (right).}\label{fig:i_fwhm_v_mag}
\end{figure}

We initially test our perturbed simulations against the UDF with a size-magnitude plot by plotting \textsc{fwhm} vs. AB magnitude.  The i-band plot is given as a representative sample, as shown in Figure $\ref{fig:i_fwhm_v_mag}$.  The
distribution should be the same for the simulations as for the original
UDF.  We use \texttt{SExtractor} parameters \textsc{fwhm\_image} and \textsc{mag\_best}.  We also test the simulations' realism through a magnitude histogram, a histogram of ellipticity components $e_1$ and $e_2$, 
and a histogram of \textsc{fwhm}. 
Since one can always increase the number of simulated galaxies in an
image, the histograms are normalized to the same area to better
observe the simulations' realism.  We define the
ellipticity components $e_1$ and $e_2$ to be

\begin{equation}
\left( e_1 \atop e_2 \right) \equiv {a^2-b^2 \over a^2+b^2} \left(\cos2\theta \atop \sin2\theta \right),
\end{equation}

\noindent where $a$, $b$, and $\theta$ are the
\texttt{SExtractor} parameters \textsc{a\_image}, \textsc{b\_image}, and \textsc{theta\_image}, namely the major and minor axes and
the angle between the major axis and the horizontal.  This convention
is generally adopted in weak lensing.

The plots discussed above reveal a strong morphological agreement between the simulated images and the originals.  
There is a good representation of objects from AB magnitude 22 to 28 for all bands up to a normalization factor.  The ellipticity histograms show a very strong representation of objects with $e_1$ and $e_2$ between $\pm 0.8$.  Beyond this, objects are not as well represented on account of the difficulty in representing highly elliptical objects as shapelets.  Lastly, the \textsc{fwhm} histogram as well shows a strong agreement across all sizes.

A more demanding test is provided by the morphological classification parameters asymmetry
(A), concentration (C), and clumpiness (S).
These morphology
characteristics have been developed in \cite{BJC,CGW,C}.  The \textit{CAS} parameters are defined in this work slightly
differently than in \cite{C}. 
We define the asymmetry, concentration, and clumpiness to be

\begin{equation}
A \equiv \frac{\sum |I_{x,y} - I_{x,y}^{180}|}{\sum |I_{x,y}|}
\end{equation}

\begin{equation}
C \equiv 5 \times \log_{10}\left(\frac{r_{80}}{r_{20}}\right)
\end{equation}

\begin{equation}
S(\sigma) \equiv 10 \times \frac{\Sigma_{xy} |I_{x,y} - I_{x,y}^{\sigma} |}{\Sigma_{xy} I_{x,y}},
\end{equation}

\noindent where $I_{x,y}$  is the flux intensity at a given pixel,
$I_{x,y}^{180}$ is the intensity at the point $180^{\circ}$ around the origin,
$r_{80}$ and $r_{20}$ are radii containing 80$\%$ and 20$\%$ of the flux
respectively, and $I_{x,y}^{\sigma}$ is the intensity after the image
is smoothed with a Gaussian kernel of width $\sigma$.  The definitions
above do not include a correction for the background as the more
typical versions do.  This is noted, but the error should average to zero
when many galaxies are used given that the noise characteristics are
the same for both the UDF and the simulated images.  We use a smoothing width $\sigma$
for $S$ to be 5 pixels.  We also use the
Petrosian Radius (R), defined to be the radius where the surface
brightness at that radius is equal to 20$\%$ of the surface brightness
integrated within that radius \cite{Petrosian}.  Massey \textit{et al} (2004)
demonstrated that the monochromatic shapelet image simulations are consistent with
real data by plotting A vs. C, A vs. S, and R vs. C for the
simulations and for the real data.  Another test is to check the mean
and RMS values for A,C, and S for the simulations against the original
UDF.  It was found that the simulations
relative to the Hubble Deep Field (HDF)
demonstrated  a roughly equal concentration while showing a lower
asymmetry and clumpiness \cite{Polar}.  Though the objects in the original
simulations from the HDF have this discrepancy,
it is relatively small, and it is concluded that the HDF simulations
are realistic \cite{MRCB}.  We demonstrate a similar recovery here.  

Examining the \textit{CAS} plots, we see generally a strong agreement between the simulated images and the
originals.   One will notice the spread in Petrosian Radius for the simulated images.  This can be explained
by the shapeletization of objects.  We have chosen to optimize the shapelet decomposition of objects to
completely model the wings of galaxies, but at the expense of sometimes truncating their central cusps. This
causes an increase in the Petrosian Radius, but should not present a large problem to methods such as weak
lensing since the shearing (and then PSF smearing) happens later; this is just a small change in the
intrinsic shape of the individual object which varies far more than the shear signal anyway.

\begin{figure}[!h]
\begin{center}
\begin{tabular}{cc}
\includegraphics[scale=0.4]{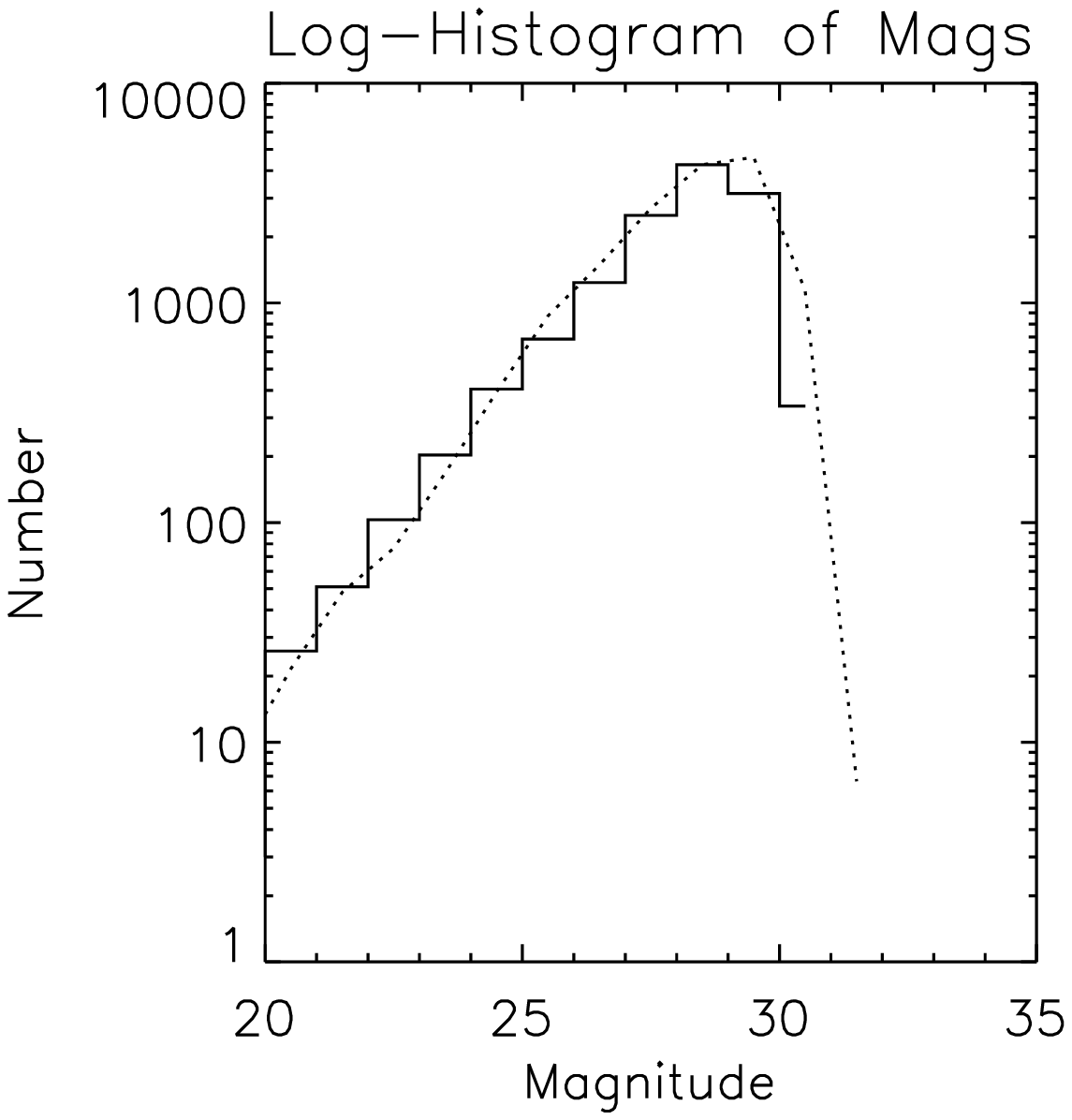}
&
\includegraphics[scale=0.4]{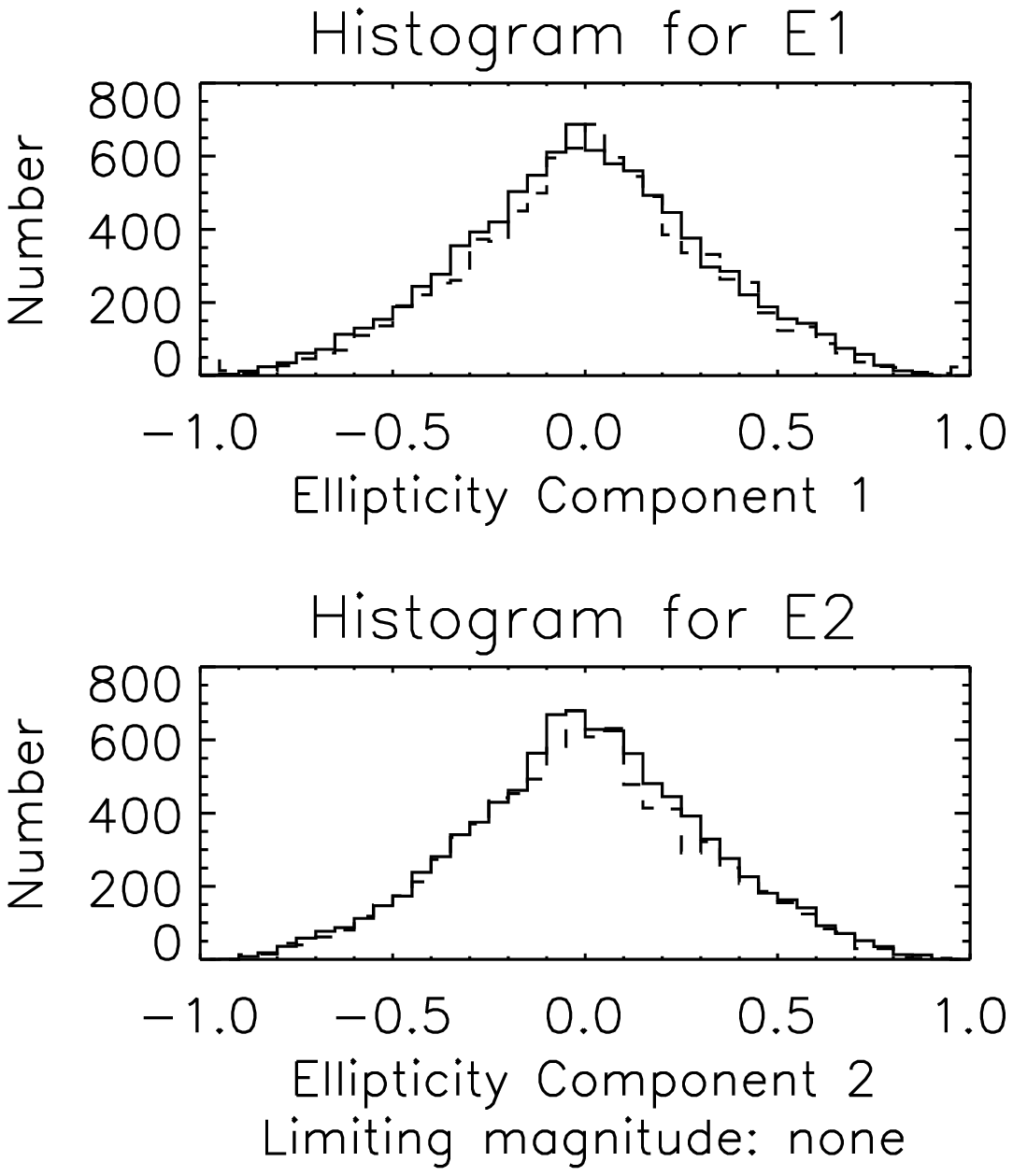}
\\
\includegraphics[scale=0.4]{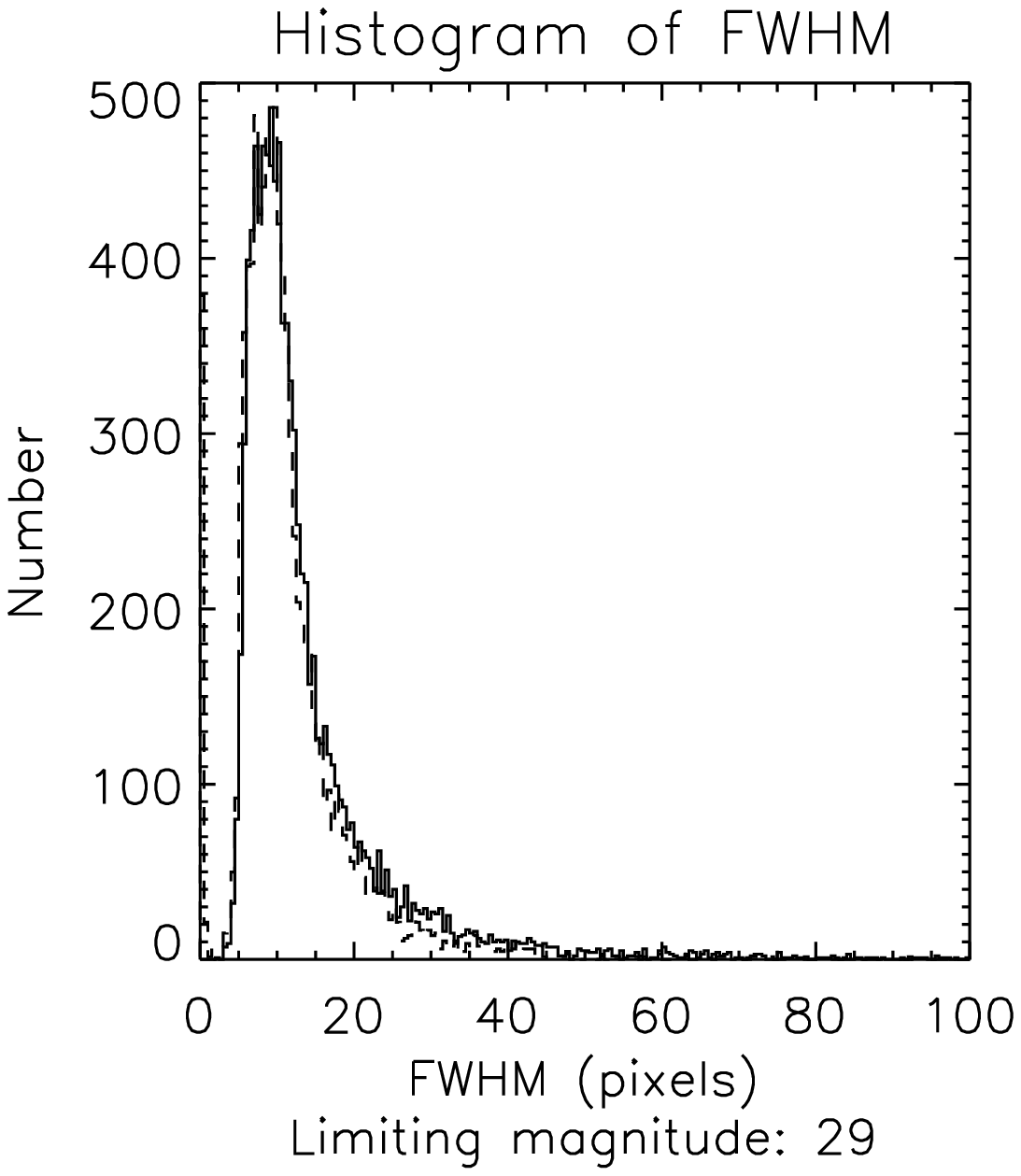}
&
\includegraphics[scale=0.4]{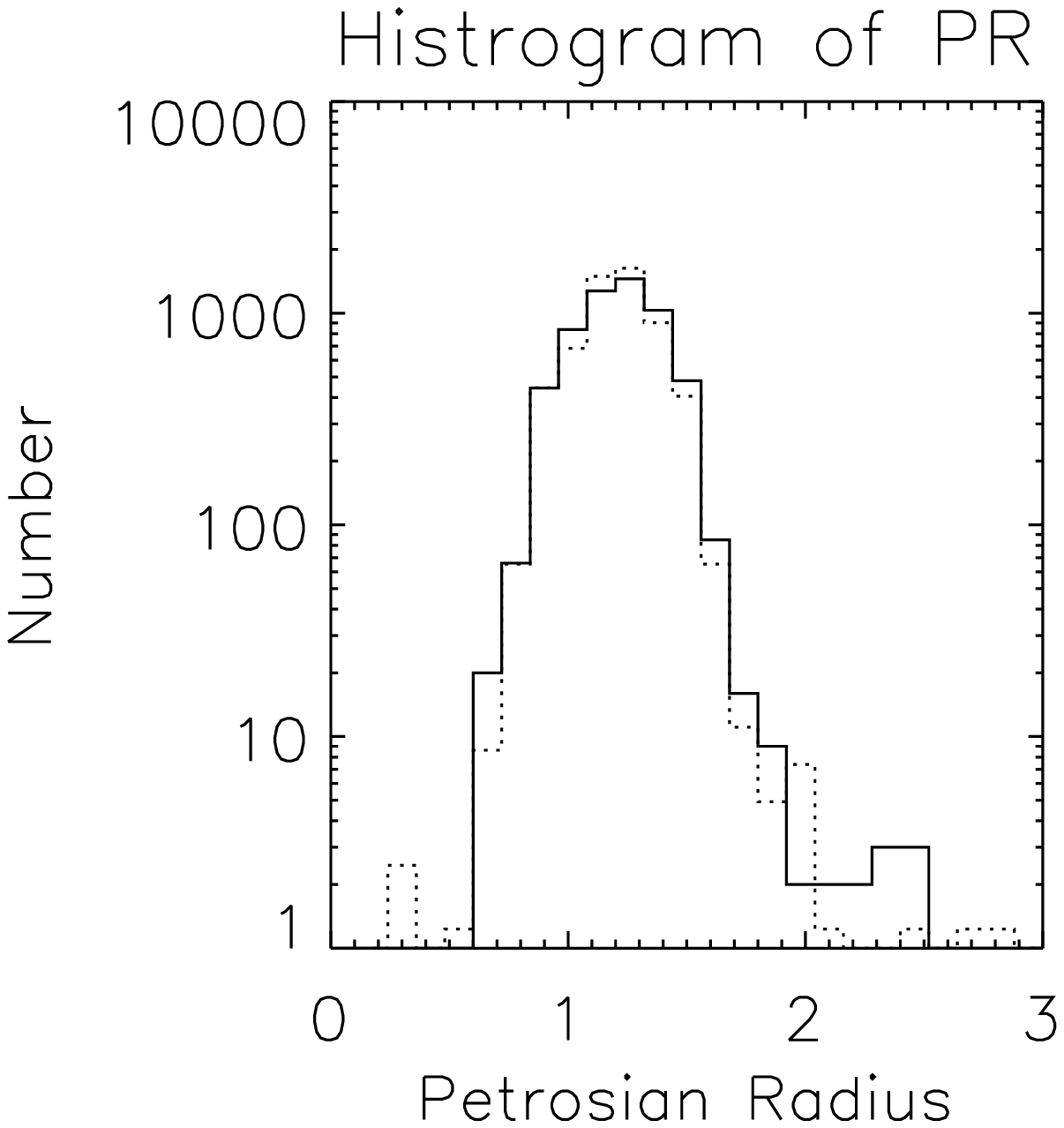}
\\
\end{tabular}
\end{center}\caption{Histograms of magnitudes (top-left),
ellipticity $e$ (top-right), \textsc{fwhm} (bottom-left), and Petrosian Radius (bottom-right) for the i-band. Dashed lines refer to the original
UDF while the solid lines refer to the simulated images.}\label{fig:i_hist}
\end{figure}

\begin{figure}[!h]
\begin{center}
\begin{tabular}{ccc}
\includegraphics[scale=0.4]{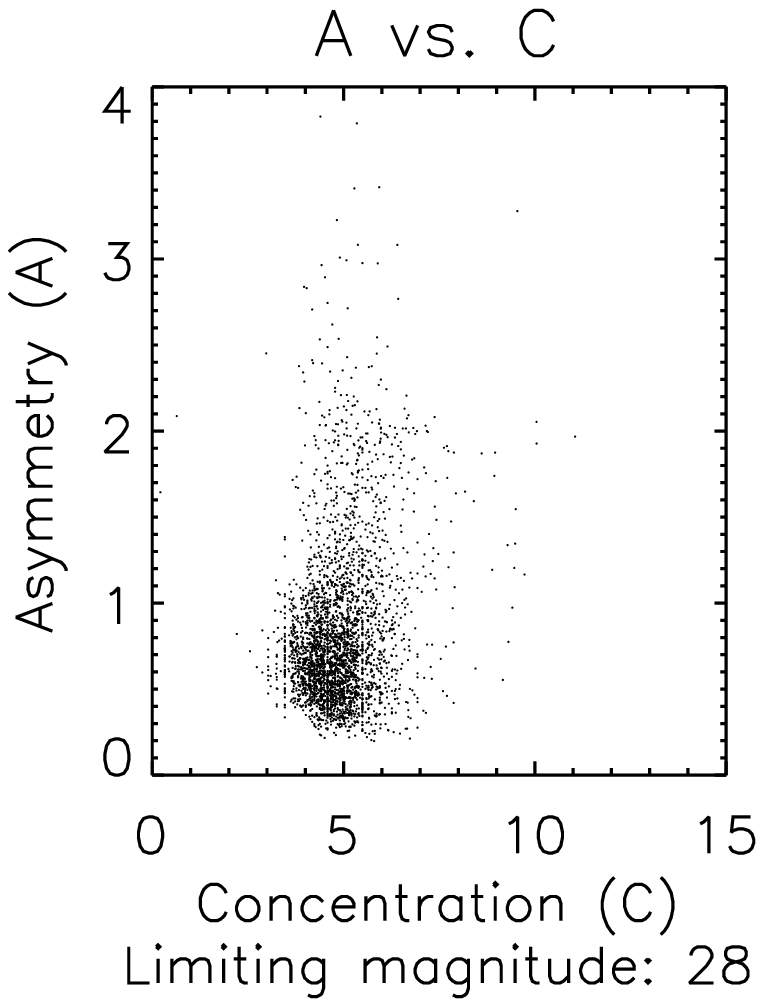}
&
\includegraphics[scale=0.4]{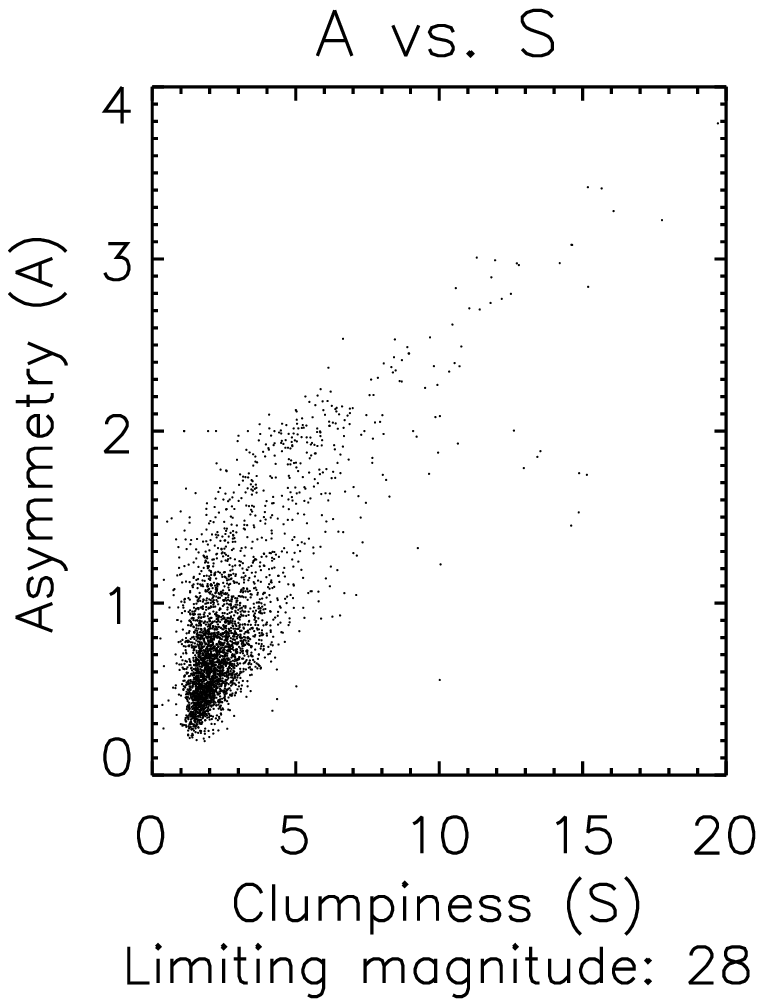}
&
\includegraphics[scale=0.4]{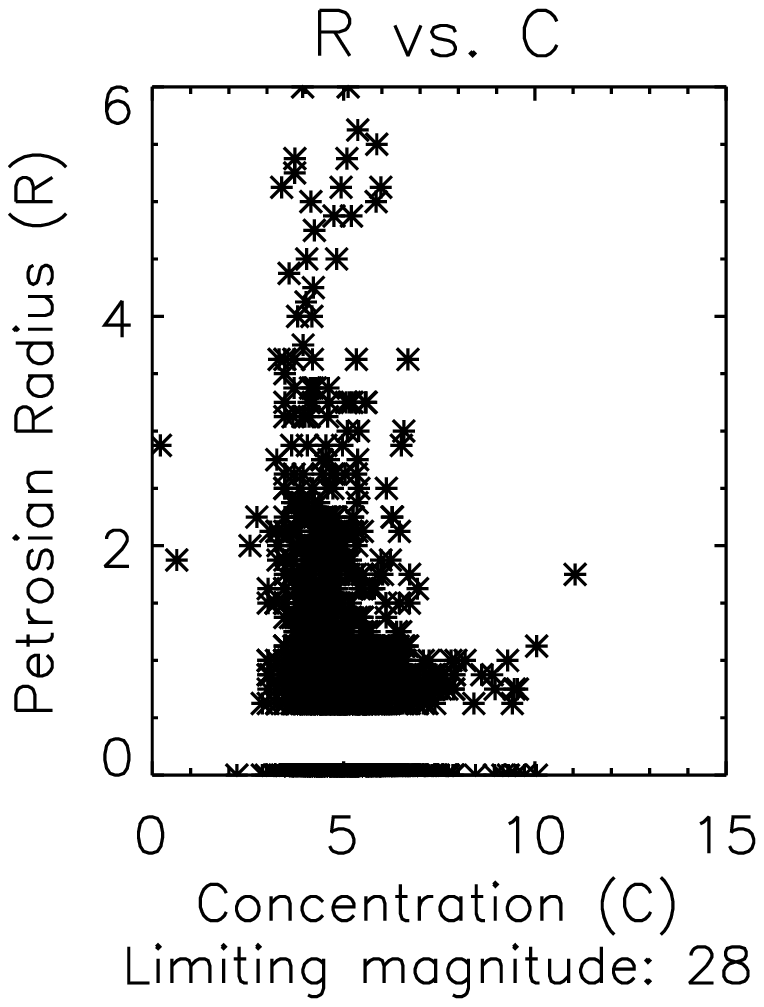}
\\
\includegraphics[scale=0.4]{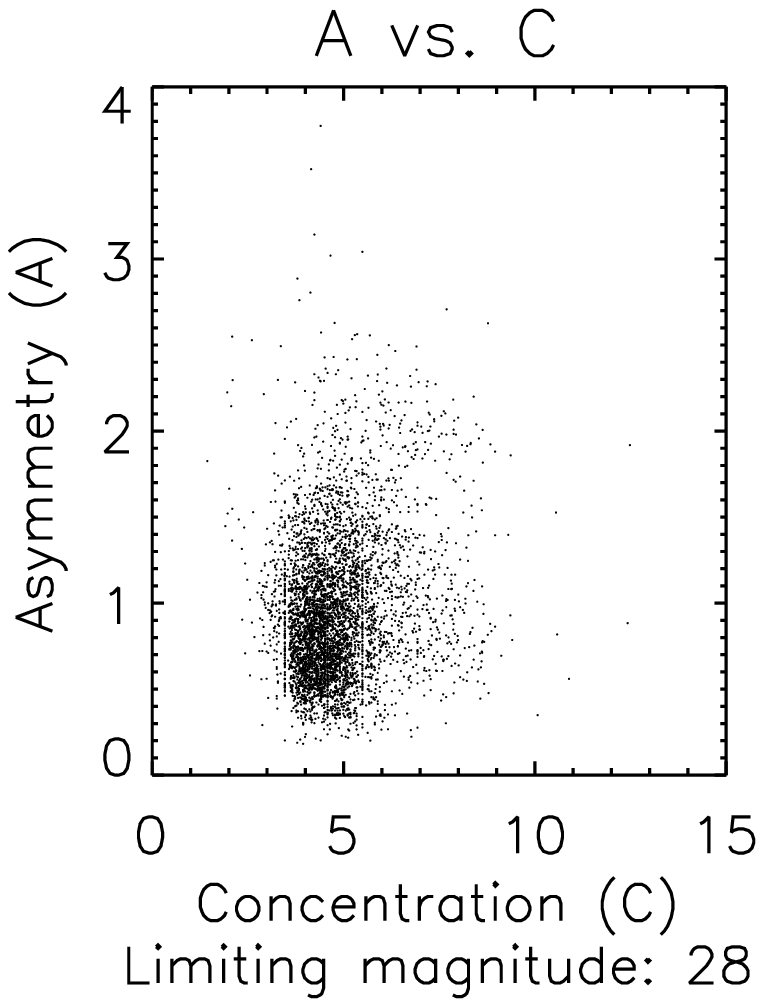}
&
\includegraphics[scale=0.4]{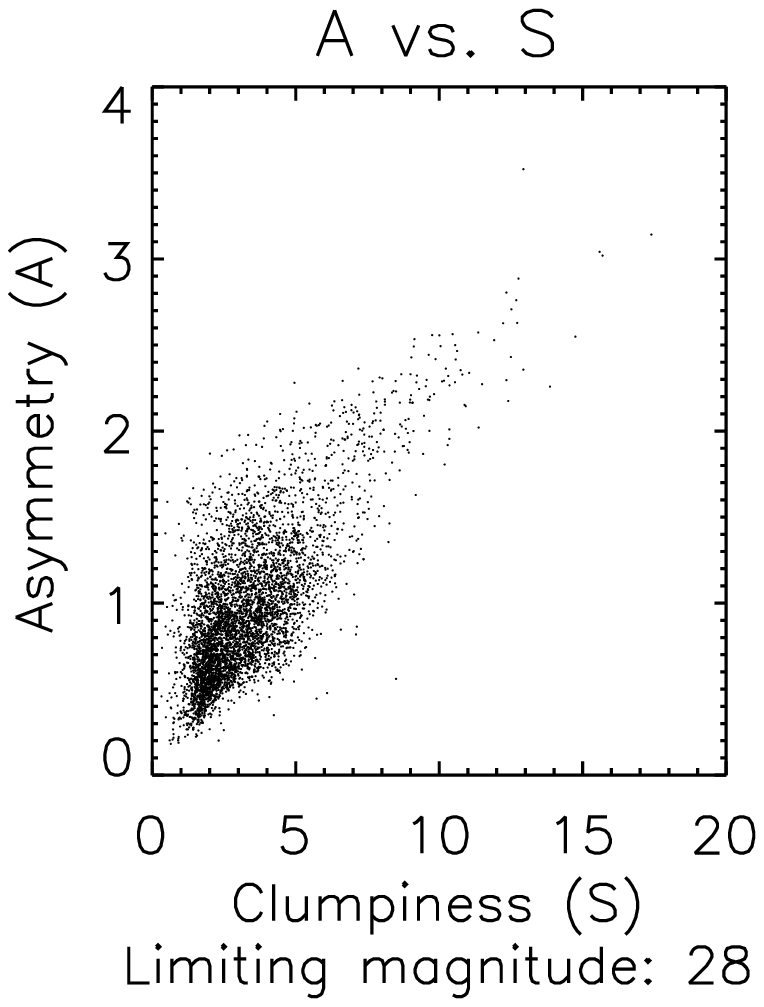}
&
\includegraphics[scale=0.4]{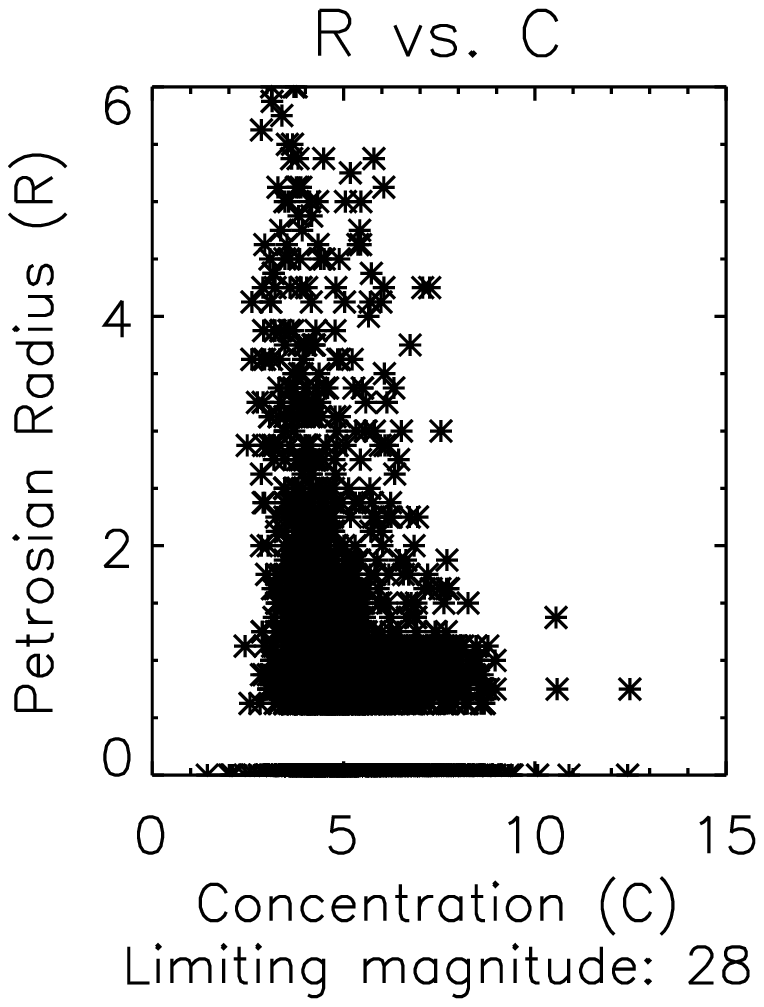}
\\
\end{tabular}
\end{center}\caption{Morphology plots for i-band real UDF (top) and
simulated images (bottom).  
}\label{fig:i_morph}
\end{figure}

In computing our results, we rejected any major outliers in the simulated images, for there were, on
occasion, hugely asymmetric objects with unrealistically high concentration and clumpiness indices.  These
objects were very large galaxies that had not properly decomposed into shapelets.  These objects were
flagged and not included in subsequent simulations.  We also rejected any asymmetry, clumpiness, or
concentration measurements in any band where a galaxy was so faint that the $CAS$ routines failed
to converge.  We also set the limiting magnitude for these statistics to be 28 as computed by
\texttt{SExtractor} as a balance between believable measurements and sufficiently deep galaxies.  The
results from the i-band simulated images are presented as a representative sample in Figures
$\ref{fig:i_fwhm_v_mag}$ through $\ref{fig:i_morph}$.   A summary of results for all bands is presented in
Table $\ref{tab:morph}$.  The overall agreement is quite good with very little deviation from the original
UDF.

\begin{figure}[!h]
\begin{center}
\begin{tabular}{cc}
\includegraphics[width=0.45\linewidth, bb=100 200 530 580, clip=]{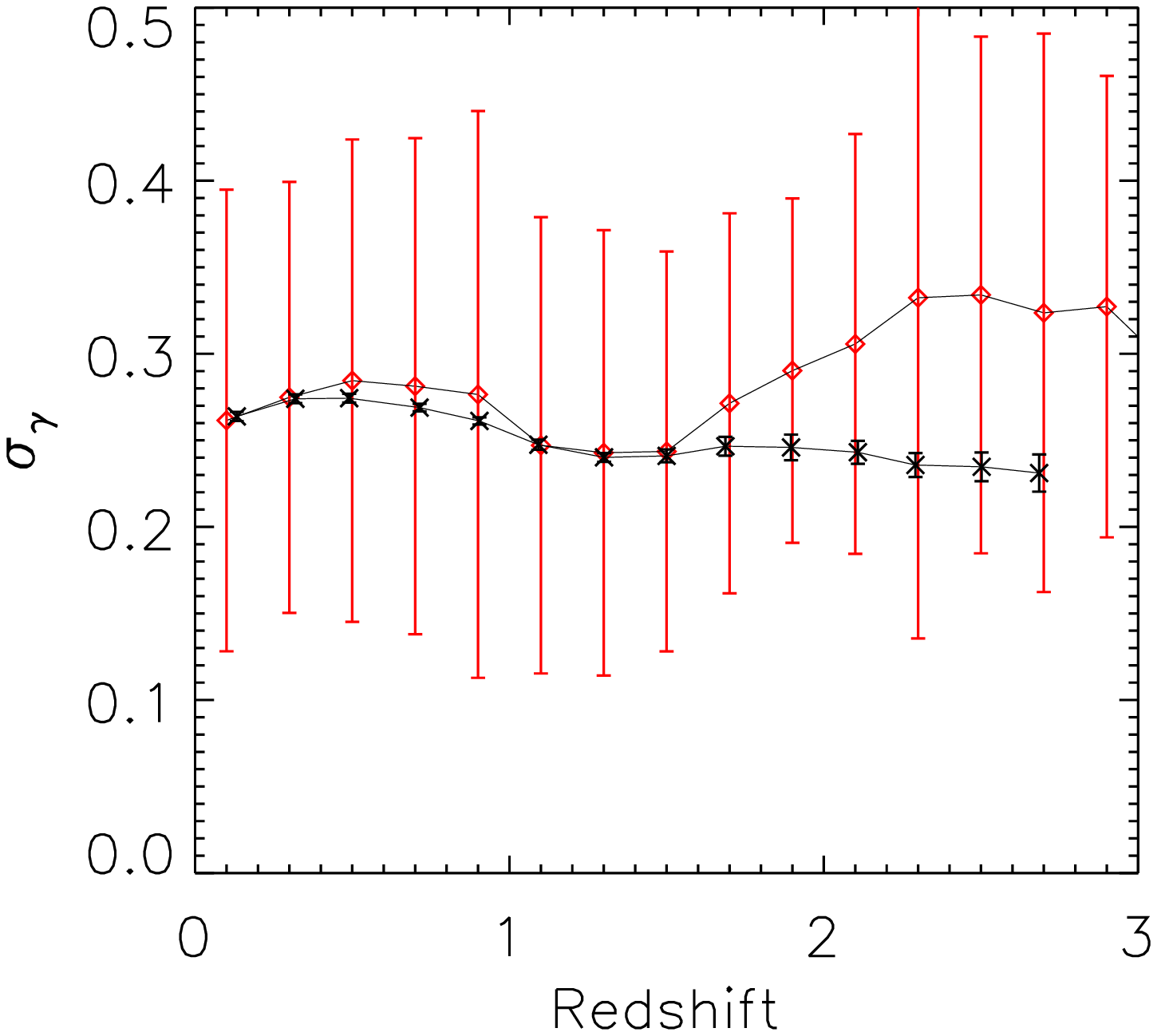}
&
\includegraphics[width=0.45\linewidth, bb=100 200 530 580, clip=]{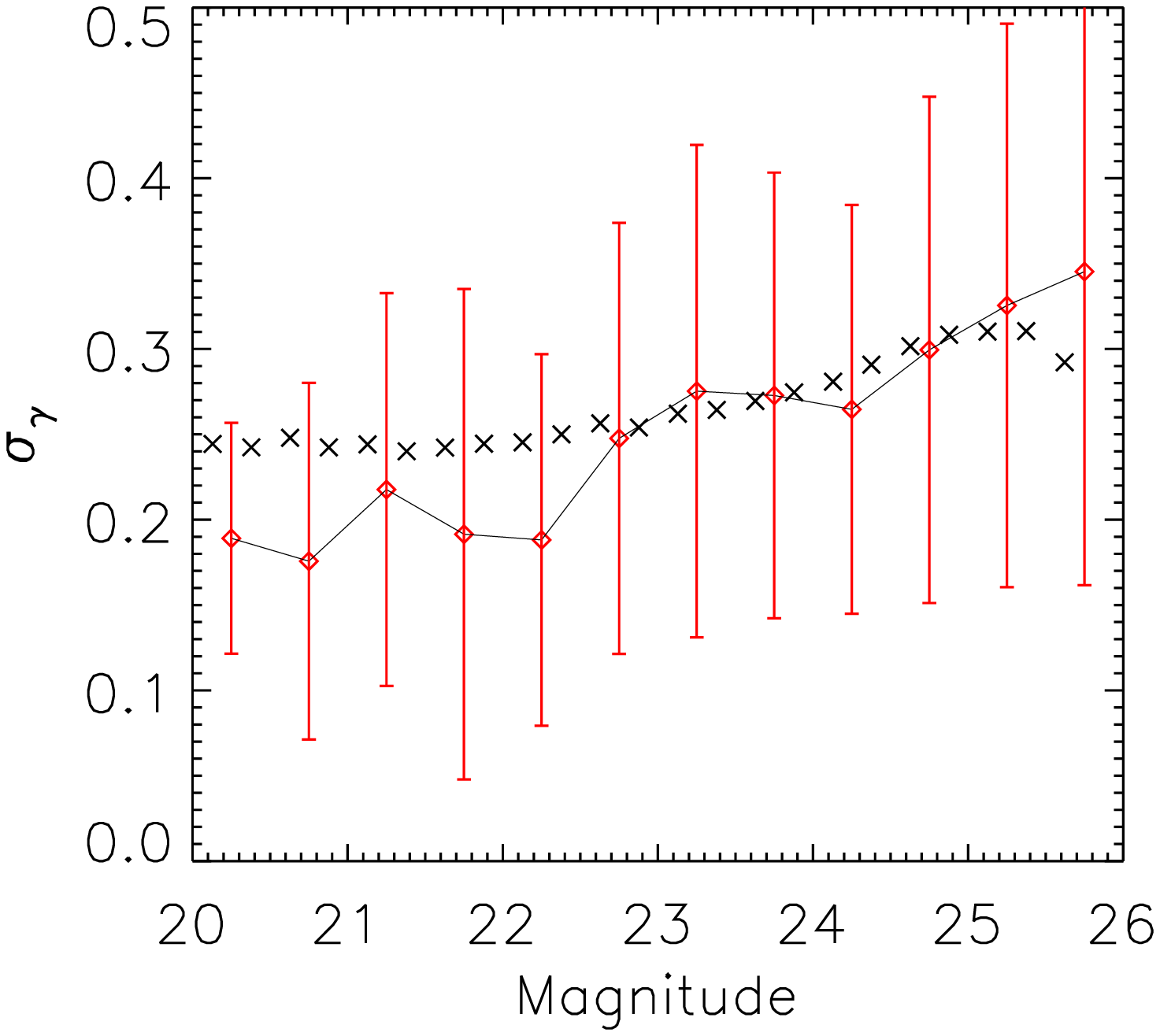}
\\
\end{tabular}
\end{center}
\caption{The scatter in measured shears, $\sigma_\gamma$, in the i-band, as a function of 
photometric redshift (left) and magnitude (right).  
$\sigma_\gamma$ is computed to be the standard deviation 
of the shear ($\gamma$), calculated in each redshift/magnitude bin.  
Red points correspond to one UDF-sized simulated image, while black crosses 
show measurements from the two square degree HST COSMOS survey \cite{Alexie}.
The simulation points are actually obtained after averaging many simulated images to reduce the error.
However, since we are drawing galaxies from a finite real population of UDF galaxies, 
the statistical error on this measurement is quickly dominated by the scatter of these galaxies,
particularly at the bright end where their numbers are limited.
We therefore represent the error bars as standard deviations rather than errors on the mean.}\label{fig:sig_gamma}
\end{figure}

As we shall use these simulations for weak lensing, of particular interest is the intrinsic ellipticity
variance ($\sigma_\gamma$), as a function of magnitude and redshift for shear-measurable galaxies.  We
include $\sigma_\gamma(z,mag)$ in Figure $\ref{fig:sig_gamma}$. We calculate the shear of a galaxy with the
weak lensing measurement method of \textit{Rhodes, Refregier, and Groth} (hereafter RRG) \cite{RRG}.  This
is a mature method developed specifically for space-based weak lensing measurements and with thorough
testing during analysis of many Hubble Space Telescope images: including the Groth Strip \cite{gs},  the
Medium Deep Survey \cite{mds},  the STIS Parallel Survey \cite{stis_ps},  and the COSMOS 2 Square Degree
Survey \cite{cosmos}. This method was also used for testing during the development of the monochromatic
shapelets image simulation pipeline described in \cite{MRCB}.  We run the RRG pipeline on the simulated
images exactly as we have run it on real HST data and make similar cuts on objects in order to obtain the
most representative results possible.

We define a shear-measurable galaxy to be one that passes several cuts. We first discard faint galaxies with
S/N less than 10, where the S/N is defined as the the ratio of the  \texttt{SExtractor} parameters
\texttt{flux\_auto} to \texttt{fluxerr\_auto}.  We also remove galaxies with ellipticity $|e|>2$, after
correction for the PSF \cite{cut}.  Note that, in the presence of image noise, especially during the PSF
correction stage, it is possible for a moment-based shape measurement method to produce  a non-physical
ellipticity $|e|>1$.  This ellipticity cut also implicitly removes galaxies for which an iterative
centroiding process in RRG failed to converge.  This includes objects for which there was a large shift away
from the initial position detected by \texttt{SExtractor}. They are usually blended objects or close pairs,
for which an accurate shape measurement would be impossible anyway.  We finally discard objects with size
$d_{RRG} = \sqrt{\frac{1}{2}(I_{xx}+I_{yy})}$ (where $I_{xx}$ and $I_{yy}$ are the weighted second order
moments) smaller than 1.2 times that of the PSF.  It is important that the cuts we make on the galaxies
useful for lensing be as realistic as possible.  Given the long history of the RRG method in space-based
weak lensing measurements we feel that running the RRG pipeline on the simulated images is the best way to
make these cuts realistic.

In Fig. $\ref{fig:sig_gamma}$, we compare the RMS shear for our simulations with real data from COSMOS \cite{Alexie}.  The high agreement within our sample error is indicative of the simulations' realism.  Running more simulations lowers the statistical error bars, but we are limited by sample variance due to the limited  number of galaxies in the UDF.  The high error bars seen in the figure reflect this uncertainty.

\begin{table}[!h]
\begin{center}
\begin{tabular}{|l|l|lllllll|}
 \hline
Filter  & Image & $<$A$>$   & rms A     & $<$C$>$   & rms C     & $<$S$>$   & rms S     & rms e \\
\hline
B   & Real UDF      & 0.82  & 0.91      & 4.66      & 4.72      & 2.83      & 3.34      & 0.45  \\
& $\delta$-function & 0.88 & 0.97 & 4.62        & 4.69     & 2.96     & 3.47     & 0.44  \\
& Perturbed & 0.92 & 1.01       & 4.60      & 4.69      & 3.28      & 3.74     & 0.44  \\
\hline
V   & Real UDF  & 0.88  & 1.02      & 4.87      & 4.94      & 2.73      & 3.46      & 0.43  \\
& $\delta$-function & 0.89  & 0.97  & 4.81      & 4.88     & 2.68      & 3.06     & 0.42  \\
& Perturbed & 0.98  & 1.07     & 4.84      & 4.95      & 3.27      & 3.74      & 0.43  \\
\hline
i   & Real UDF  & 0.82  & 1.03      & 4.88      & 4.96      & 2.63      & 3.45      & 0.42  \\
& $\delta$-function & 0.88 & 0.97       & 4.82      & 4.88      & 2.77      & 3.11      & 0.42  \\
& Perturbed & 0.95  & 1.04      & 4.81      & 4.93      & 3.27      & 3.70     & 0.43  \\
\hline
z   & Real UDF  & 0.75  & 0.87      & 4.72      & 4.80      & 2.76      & 3.19      & 0.43  \\
& $\delta$-function & 0.88 & 0.95       & 4.80      & 4.91      & 3.05      & 3.39     & 0.41  \\
& Perturbed     & 0.90  & 1.03      & 4.70     & 4.82      & 3.17      & 3.53      & 0.43  \\
\hline
\end{tabular}
\end{center}\caption{Final results table comparing morphologies of the
UDF with the simulated images.  The $\delta$-function images refer to simulated images created without perturbing the shapelet coefficients ($\lambda_i$).  The perturbed images were created by the smoothing method discussed in \S $\ref{sec:im_create}$.  For all the images, the limiting AB magnitude was 28.}\label{tab:morph}
\end{table}

\section{Conclusions}

We presented a method to create an arbitrary amount of
3-dimensional, color,
simulated, unique deep space images.  The simulations are created by
perturbing a galaxy's polar shapelet coefficients in such a way as to
create unique but realistic objects.  The previous simulation pipeline has been expanded 
to correlate
morphologies across four wavelength bands and now also include a
redshift distribution.  Though we currently use four wavelength bands, our simulation pipeline is flexible enough to include an arbitrary number of colors, should such a data set become available and useful.

Our simulations were tested by comparing them to the
original UDF images.  They were found to have similar morphologies
to the original galaxies.  Additionally, the weak lensing cosmological properties of the simulations were tested against COSMOS HST data.  Within reasonable error, our simulations were found to be consistent.

\section*{Acknowledgements}

The authors thank the Caltech SURF program, the Berkeley Physics
Undergraduate Research Scholars program, and the Jet Propulsion Laboratory, run under a contract with NASA by Caltech, for their support. 
We would also like to thank Richard Ellis for further support and
enthusiasm regarding the project and Alexandre Refregier for his ideas and enthusiasm.  The project would not have gotten
started on the right path were it not for Will High's help.  Thanks
also to Peter Capak for useful insights in simultaneous detection with
\texttt{SExtractor}.  Dave Johnston provided the PSF for SNAP and was helpful in calibrating the measurements of $\sigma_\gamma$.  Thanks also to Alexie Leauthaud for help in shear measurement.  JR and MF were supported in part by NASA grant BEFS-399131.02.02.01.07.

\end{document}